\documentclass[prl,floatfix,showpacs,twocolumn,preprintnumbers,amsmath,amssymb,superscriptaddress]{revtex4}
\usepackage[sort&compress]{natbib}
\usepackage{graphicx,float,datetime}
\usepackage{hyperref,wasysym}
\usepackage{epstopdf,subfigure}
\usepackage{enumitem}
\usepackage{bm}
\usepackage[utf8]{inputenc}

\newcommand\tauem{\overset{\textit{\tiny e-m}}{\mathcal{T}}}
\newcommand\numberthis{\addtocounter{equation}{1}\tag{\theequation}}

\newcommand*\xbar[1]{%
  \hbox{%
    \vbox{%
      \hrule height 0.5pt % The actual bar
      \kern0.5ex%         % Distance between bar and symbol
      \hbox{%
        \kern-0.1em%      % Shortening on the left side
        \ensuremath{#1}%
        \kern-0.1em%      % Shortening on the right side
      }%
    }%
  }%
} 

\newcommand{\udt}[3]{#1^{#2}_{\phantom{#2}#3}}

\def\ber{\begin{eqnarray}}
\def\eer{\end{eqnarray}}
\def\beq{\begin{equation}}
\def\eeq{\end{equation}}

\newdateformat{mydate}{\THEDAY\hspace{3pt}\monthname[\THEMONTH] \THEYEAR}

\begin{document}

\begin{center}
\title{Quark Stars in $f(T, \mathcal{T})-$Gravity}
\date{\mydate\today}
\author{Mark Pace\footnote{mark.pace.10@um.edu.mt}}
\affiliation{Department of Physics, University of Malta, Msida, MSD 2080, Malta}
\affiliation{Institute of Space Sciences and Astronomy, University of Malta, Msida, MSD 2080, Malta}
\author{Jackson Levi Said\footnote{jackson.said@um.edu.mt}}
\affiliation{Department of Physics, University of Malta, Msida, MSD 2080, Malta}
\affiliation{Institute of Space Sciences and Astronomy, University of Malta, Msida, MSD 2080, Malta}

\begin{abstract}
{
\noindent
We derive a working model for the Tolman-Oppenheimer-Volkoff equation for quark star systems within the modified $f(T, \mathcal{T})$-gravity class of models. We consider $f(T, \mathcal{T})$-gravity for a static spherically symmetric space-time. In this instance the metric is built from a more fundamental tetrad vierbein from which the metric tensor can be derived. We impose a linear $f(T)$ parameter parameter, namely taking $f=\alpha T(r) + \beta \mathcal{T}(r) + \varphi$ and investigate the behavior of a linear energy-momentum tensor trace, $\mathcal{T}$. We also outline the restrictions which modified $f(T, \mathcal{T})$-gravity imposes upon the coupling parameters. Finally we incorporate the MIT bag model in order to derive the mass-radius and mass-central density relations of the quark star within $f(T, \mathcal{T})$-gravity.
}
\end{abstract}


\maketitle

\end{center}

%%------------------------Section-------------------------
\section{I. Introduction}\label{sec:intro}
\noindent
In recent years it has been shown that the Universe is accelerating in its expansion \cite{garnavich1998supernova, riess1998observational}. In order to explain this one can introduce the concept of the cosmological constant \cite{copeland2006dynamics, bamba2012dark}. Together with the inclusion of dark matter we get the $\Lambda$CDM model which explains a whole host of phenomena within the universe. \cite{feng2005dark, guo2005cosmological, boehmer2011existence}. Another approach to explaining this acceleration is to modify the gravitational theory itself with alternative theories of gravity an example of which is $f(R)$-gravity \cite{de2010f, Capozziello:2011et, Cai:2015emx, Capozziello:2009nq}.
\\ 
\\
$f(T)$-gravity uses a ``teleparallel'' equivalent of GR (TEGR) \cite{iorio2012solar} approach, in which instead of the torsion-less Levi-Civita connection, the Weitzenb\"{o}ck connection is used, with the dynamical objects being four linearly independent vierbeins \cite{unzicker2005translation, hayashi1979new}. The Weitzenb\"{o}ck connection is curvature-free and describes the torsion of a manifold. In the current case we consider a pure tetrad \cite{Tamanini:2012hg}, meaning that the torsion tensor is formed by a multiple of the tetrad and its first derivative only. The Lagrangian density can be constructed from this torsion tensor under the assumption of invariance under general coordinate transformations, global Lorentz transformations, and the parity operation \cite{Tamanini:2012hg, iorio2012solar, hayashi1979new, Capozziello:2009nq}. Also the Lagrangian density is second order in the torsion tensor \cite{iorio2012solar, hayashi1979new}. Thus $f(T)$-gravity generalises the above TEGR formalism, making the gravitational lagrangian a function of $T$ \cite{iorio2012solar, Cai:2015emx, Capozziello:2009nq}.
\\
\\
Our study involves deriving a working model for the TOV equation within a new modification of $f(T)$ class gravity, namely $f(T, \mathcal{T})$-gravity. 
%Such a theory allows for a general coupling of the torsion scalar $T$ with the trace of the matter energy momentum tensor $\mathcal{T}$ \cite{Cai:2015emx, Planck:2013jfk}. 
There is no theoretical reason against couplings between the gravitational sector and the standard matter one \cite{Cai:2015emx}. $f(T, \mathcal{T})$-gravity takes inspiration from $f(R, \mathcal{T})$-gravity \cite{Cai:2015emx, Planck:2013jfk} where instead of having the Ricci scalar coupled with the trace of the energy momentum tensor $\mathcal{T}$, one couples the torsion scalar $T$ with the trace of the matter energy-momentum tensor $\mathcal{T}$ \cite{Cai:2015emx, Planck:2013jfk, Capozziello:2009nq}.
\\
\\
Recently a modification to this theory has been propose, that of allowing for a general functional dependence on the energy momentum trace scalar, $\udt{\mathcal{T}}{\mu}{\mu}=\mathcal{T}$. 
\\
\\
%Here the gravitational lagrangian takes the form of an arbitrary function $f(T,\mathcal{T})$ \cite{harko2014f, momeni2014cosmological}. The energy momentum tensor may allow for interesting behaviours in terms of the cosmological constant \cite{nassur2015early, Starobinsky:1999yw, Harko:2011kv}. $f(T,\mathcal{T})$
Our interest is in studying the behaviour of spherically symmetric compact objects in this theory with a specific linear function being considered, namely $f(T,\mathcal{T})=\alpha T(r) + \beta\mathcal{T}(r) + \varphi$ where $\alpha,\,\beta$ are arbitrary constants, and $\varphi$ we take as the cosmological constant. We consider the linear modification since it is the natural first functional form to consider, and the right place to start to understand how the trace of the stress-energy tensor might effect $f(T,\mathcal{T})$ gravity. In particular, our focus is on quark stars in $f(T,\mathcal{T})$ gravity. Besides the possibility of the existence of these exotic stars, this is also a good place to study the behavior of modified theories of gravity in terms of constraints. Moreover, this also opens the door to considerations of stiff matter in early phase transitions \cite{astashenok2015nonperturbative}.
\\
\\
The plan of this paper is as follows; In section 2 we outline the theoretical background of the model. 
%Throughout this paper we take the vierbein field as $e_{A} \left(x^{\mu} \right)$ which forms the dynamical variable of TEGR and its $f(T, \mathcal{T})$ extension \cite{iorio2012solar}. This forms an orthonormal basis for the tangent space at each point $x^{\mu}$ of the manifold, i.e. $\textbf{e}_{A} \cdot \textbf{e}_{B} = \eta_{AB}$, where $\eta_{AB} = \text{diag}(1, -1, -1, -1)$. The vector $\textbf{e}_{A}$ can be analysed with the use of its components $e^{\mu}_{\enspace A}$ in a coordinate basis, that is $\textbf{e}_{A} = e^{\mu}_{\enspace A} \partial_{\mu}$ \cite{farrugia2016solar, paliathanasis2016cosmological}. 
In section 3 we consider the rotated tetrad and use this to derive the TOV equation in $f(T, \mathcal{T})$-gravity in section 4. Section 5 will then present the contrasting mass-radius relations derived using the MIT bag model, which we derive numerically. Finally in section 6 we discuss the results.

%------------------------Section-------------------------
\section{II. Field Equations of $f(T, \mathcal{T})$-gravity}\label{sec:eqfttgrav}
%------------------------Section-------------------------
\noindent
The concept of $f(T, \mathcal{T})$-gravity is a generalisation of $f(T)$-gravity and thus based on the Weitzenbock's geometry. We will follow a similar notation style as that given in \cite{iorio2012solar, farrugia2016solar, paliathanasis2016cosmological, Capozziello:2011et, Cai:2015emx}. Using: greek indices $\mu, \nu, \dots$ and capital Latin indices $A, B, \dots$ over all general coordinate and inertial coordinate labels respectively. While lower case Latin indices i.e. $i, j, \dots$ and $a, b, \dots$ cover spatial and tangent space coordinates 1, 2, 3, respectively \cite{farrugia2016solar, paliathanasis2016cosmological, Capozziello:2011et, Cai:2015emx}.
\\
\\
%The curvatureless Weitzenb\"{o}ck connection \cite{cooney2010neutron, nassur2015early} is given by
%
%\begin{equation}
%\Gamma^{\rho}_{\enspace \mu \nu} = e^{\rho}_{i} \partial_{\nu} e^{i}_{\mu} + \omega^{i}_{\enspace \lambda \nu} e^{i}_{\mu}.
%\end{equation}
\noindent
The non-vanishing torsion \cite{farrugia2016solar, paliathanasis2016cosmological, Krssak:2015oua} is given by
\\
\begin{multline}
T^{\lambda}_{\enspace \mu \nu} \left( e^{\lambda}_{\enspace \mu}, \omega^{\lambda}_{\enspace i \mu} \right) = \partial_{\mu} e^{\lambda}_{\enspace \nu} - \partial_{\nu} e^{\lambda}_{\enspace \mu} + \\ \omega^{\lambda}_{\enspace i \mu} e^{i}_{\enspace \nu} - \omega^{i}_{\enspace \lambda \nu} e^{i}_{\mu}.
\end{multline}
\noindent
In TEGR one uses the teleparallel spin connection, which by construction gives vanishing curvature, thus all the information of the gravitational field is embedded in the torsion tensor, while the gravitational Lagrangian is the torsion scalar \cite{Krssak:2015oua}. The contorsion tensor is then defined as

\begin{equation}
K^{\mu \nu}_{\enspace \enspace \rho} = - \dfrac{1}{2} \left(T^{\mu \nu}_{\enspace \enspace \rho} - T^{\nu \mu}_{\enspace \enspace \rho} - T_{\rho}^{\enspace \mu \nu} \right),
\end{equation}
\noindent
while the superpotential of teleparallel gravity is defined by \cite{farrugia2016solar, paliathanasis2016cosmological}

\begin{equation}
S_{\rho}^{\enspace \mu \nu} = \dfrac{1}{2} \left( K^{\mu \nu}_{\enspace \enspace \rho} + \delta^{\mu}_{\rho} T^{\alpha \nu}_{\enspace \enspace \alpha} - \delta^{\nu}_{\rho} T^{\alpha \mu}_{\enspace \enspace \alpha} \right).
\end{equation}
\noindent
The torsion scalar \cite{farrugia2016solar, paliathanasis2016cosmological} is then given as

\begin{equation} \label{Torsion_scalar_def}
T = S_{\rho}^{\enspace \mu \nu} T^{\rho}_{\enspace \mu \nu}.
\end{equation} 
\noindent
As in the analogous $f(R,T)$ theories \cite{harko2011f}, the gravitational lagrangian is generalized to $f(T,\mathcal{T})$ giving \cite{nassur2015early, harko2014f}

\begin{equation}
\label{action}
S = - \dfrac{1}{16 \pi G} \int d^{4} x e \left[ f\left( T, \mathcal{T} \right) + \mathcal{L}_{m}  \right],
\end{equation}
where $\mathcal{T} = \delta^{\nu}_{\mu}\mathcal{T}_{\nu}^{\;\mu}$ and is the trace of the energy-momentum tensor while $\mathcal{L}_m$ is the matter Lagrangian density \cite{nassur2015early}. In this instance $f$ is an arbitrary function of the torsion scalar $T$ and the trace of the energy-momentum tensor $\mathcal{T}$ \cite{nassur2015early}. The variation of the action defined in Eq.(\ref{action}) with respect to the tetrad leads to the field equations

\begin{widetext}
\begin{multline}
\label{fieldequations}
e^{\rho}_{i} S_{\rho}^{\; \mu \nu} \partial_{\mu} T f_{TT} + e^{\rho}_{i} S_{\rho}^{\; \mu \nu} f_{T \mathcal{T}} \mathcal{T} + e^{-1} \partial_{\mu} \left( e e^{\rho}_{i} S_{\rho}^{\; \mu \nu} \right) f_{T} \\ + e^{\mu}_{i} T^{\lambda}_{\; \mu \kappa} S_{\lambda}^{\nu \kappa} f_{T} - \dfrac{e^{\nu}_{i}f}{4} + f_{T} \omega^{i}_{\enspace \lambda \nu} S_{i}^{\enspace \nu \mu} - \dfrac{f_{\mathcal{T}}}{2} \left( e^{\lambda}_{i}  \mathcal{T}_{\lambda}^{\;\nu} + p(r) e^{\nu}_{i} \right) = - 4 \pi e^{\lambda}_{i} \tauem^{\; \nu}_{\lambda},
\end{multline}
\end{widetext}
\noindent
where $f_{\mathcal{T}} = \dfrac{\partial f}{\partial \mathcal{T}}$, and $ f_{T \mathcal{T}} = \dfrac{\partial^2 f}{\partial T \partial \mathcal{T}}.$
In our case we take the spin connection as $\omega^{i}_{\enspace \lambda \nu} = 0$ from the start \cite{Krssak:2015oua, Bengochea:2008gz, Ferraro:2006jd, Ferraro:2008ey, Linder:2010py}.

\section{III. Rotated Tetrads in $f(T, \mathcal{T})$-gravity}
\noindent
We take a spherically symmetric metric for our system which has a diagonal structure \cite{deliduman2011absence}

\begin{equation} \label{metric}
ds^2 = - e^{A(r)}dt^2 + e^{B(r)}dr^2 + r^2 d \theta^2 + r^2 \sin^2 \theta d \phi^2,
\end{equation}

\noindent
and consider the fluid inside the star to be that of a perfect fluid which yields a diagonal energy-momentum tensor of

\begin{equation}
\tauem^{\; \nu}_{\lambda} = diag ( - \rho(r), p(r), p(r), p(r)),
\end{equation}

\noindent
where $\rho(r)$ and $p(r)$ are the energy density and pressure of the fluid respectively and the time dependence will be suppressed for brevity \cite{deliduman2011absence}. These also make up the matter functions which, along with the metric functions, $A(r)$ and $B(r)$, are also taken to be independent of time. Thus the system is taken to be in equilibrium \cite{boehmer2011existence, deliduman2011absence}. The equation of conservation of energy is given by

\begin{equation} \label{conservation}
\dfrac{dp(r)}{dr} = - (\rho(r) + p(r)) \dfrac{dA(r)}{dr}.
\end{equation}

\noindent
As is in \cite{Tamanini:2012hg} we use the following rotated tetrad

\begin{widetext}
\[
e_{\mu}^{a}   =
 \left( \begin{array}{cccc}
e^{\dfrac{A(r)}{2}} & 0 &  0 & 0 \\
0 & e^{\dfrac{B(r)}{2}} \sin{\theta} \cos{\phi} & e^{\dfrac{B(r)}{2}} \sin{\theta} \sin{\phi} & e^{\dfrac{B(r)}{2}} \cos{\theta} \\
0 & -r \cos{\theta} \cos{\phi} & -r \cos{\theta} \sin{\phi} & r \sin{\theta}\\
0 & r \sin{\theta} \sin{\phi} & - \sin{\theta} \cos{\phi} & 0 \end{array} \right)\]
\end{widetext}

\noindent
We take this form of vierbein because it gives us more degrees of freedom \cite{Faraoni:2000wk} and it allows us to obtain a static and spherically symmetric wormhole solution in our standard formulation of $f(T, \mathcal{T})$-gravity \cite{Faraoni:2000wk, Paliathanasis:2014iva}.
\\
\\
Moreover, this particular form of the tetrad is what's called a pure tetrad \cite{Krssak:2015oua}. This means that the spin connection elements of this tetrad vanish and the ensuing field equations do not need to consider spin connection terms \cite{Krssak:2015oua}.
\\
\\
Inserting this vierbein into the field equations, from Eq.(\ref{Torsion_scalar_def}) we get the resulting torsion scalar

\begin{equation} \label{torsionscalar}
T(r) = \dfrac{2 e^{-B(r)}}{r^2} \left(1-e^{\dfrac{B(r)}{2}} \right) \left(1 - e^{\dfrac{B(r)}{2}} + r A'(r) \right),
\end{equation}

\noindent
where the prime denotes derivative with respect to $r$. The $f(T, \mathcal{T})$ field equations result in five independent relations

\begin{widetext}
\begin{multline}
\label{equation00}
4 \pi \rho(r) = \dfrac{f}{4} + \dfrac{e^{\dfrac{-B(r)}{2}} f_{T}}{2 r^2} \left(-2 + 2 e^{\dfrac{B(r)}{2}} + r A'(r) \left(-1 + e^{\dfrac{B(r)}{2}} \right) + r B'(r) \right) \\ + \dfrac{f_{\mathcal{T}}}{2} \left( \rho(r) - p(r) \right) + \dfrac{e^{-B(r)}f_{TT}T'(r)}{r} \left(-1 + e^{\dfrac{B(r)}{2}} \right) + \dfrac{e^{-B(r)}f_{T\mathcal{T}}\mathcal{T}'(r)}{r}\left(-1 + e^{\dfrac{B(r)}{2}} \right),
\end{multline}

\begin{multline}
\label{equation11}
4 \pi p(r) = \dfrac{f}{4} + \dfrac{e^{\dfrac{-B(r)}{2}} f_{T}}{2 r^2} \left(2 \left(-1 + e^{\dfrac{B(r)}{2}} \right) + \left(-2 + e^{\dfrac{B(r)}{2}} \right) r A'(r) \right) - f_{\mathcal{T}} p(r).
\end{multline}
\end{widetext}

\noindent
For the only non-vanishing non-diagonal element $(i=1,\, \nu=2)$

\begin{equation}
\label{equation12}
e^{\dfrac{-B(r)}{2}} \dfrac{\cot{\theta}}{2r^2} \left( f_{TT} T'(r) + f_{T \mathcal{T}} \mathcal{T}'(r) \right) = 0
\end{equation}

\noindent
Together these equations govern the behaviour of the compact star.

\section{IV. TOV equations in $f(T, \mathcal{T})$-gravity}
\noindent
We take $f = \alpha T(r) + \beta \mathcal{T}(r) + \varphi$ as our lagrangian function where $\mathcal{T}(r) = \rho(r) - 3p(r)$ \cite{harko2014f}. Consider Eq.(\ref{equation11}), and solving for $A'(r)$ we find

\begin{align*} \label{Adash}
A'(r) = e^{B(r)} \bigg( \dfrac{1}{r} - \dfrac{e^{-B(r)}}{r} - \dfrac{8 \pi p(r) r}{\alpha} + \dfrac{\varphi r}{\alpha} \\ + \dfrac{\beta r}{2 \alpha} \left(\rho(r) - 7 p(r) \right) \bigg).
\numberthis
\end{align*}

\noindent
Substituting this into Eq.(\ref{equation00}) and reducing yields

\begin{multline}
r B'(r) e^{-B(r)} - e^{-B(r)} = \dfrac{8 \pi \rho(r) r^2}{\alpha} - \dfrac{\varphi r^2}{2 \alpha} \\ - \dfrac{\beta r^2}{2 \alpha} \left(3 \rho(r) - 5 p(r) \right) - 1.
\end{multline}

\noindent
We invoke the density equation for an inhomogeneous body
\begin{equation} \label{densityeq}
M(r) = \int^{r}_{0} 4 \pi r^2 \rho(r) dr,
\end{equation}
\noindent
We invoke the MIT bag model \cite{Kpadonou:2015eza} since it represents the EoS of quark stars 

\begin{equation} \label{MITbag}
p(r) = \omega \left( \rho(r) - 4 \gamma_0 \right).
\end{equation}
\noindent
We thus get

\begin{equation}\label{modifedschawz}
e^{-B(r)} = 1 + \dfrac{M(r)}{\alpha r} \Sigma + \dfrac{r^{2}}{3 \alpha} \xi,
\end{equation}
\noindent
where $\Sigma =  -2 + \dfrac{3 \beta}{8 \pi} - \dfrac{5 \beta \omega}{8 \pi} $ and $\xi = \dfrac{\varphi}{2} + 10 \beta \gamma_0$.
Substituting this into Eq.(\ref{Adash}) and then invoking the conservation Eq.(\ref{conservation}) we get a relation between pressure, $p(r)$, and radius, $r$, in this form

\begin{widetext}
\begin{multline} \label{TOV_dp}
\dfrac{dp(r)}{dr} = \dfrac{(\rho(r) + p(r))}{\alpha} \left( 
8 \pi p(r) r + \dfrac{M(r)}{r^2}\Sigma - \dfrac{\xi}{2}(\rho(r) - 7 p(r)) - \dfrac{5 r \varphi}{6} + \dfrac{10 \beta \gamma_0 r}{3} \right) \\ \left( 1 + \dfrac{M(r)}{\alpha r} \Sigma + \dfrac{r^{2}}{3 \alpha} \xi \right)^{-1}.
\end{multline}
\end{widetext}
\noindent
The mass-radius relation is also derived \cite{astashenok2013further, astashenok2015extreme}, using our modified Schwarzschild solution found in Eq.(\ref{modifedschawz}) and we get the following result

\begin{align*} \label{TOV_dm}
\dfrac{dM(r)}{dr} = 4 \pi r^2 \bigg( 5 p(r) \beta + 16 \pi \rho(r) - 3 \beta \rho(r) \\ - 20 \beta \omega \gamma_{0} \bigg)  \bigg( 16 \pi + \beta  \left( 5 \omega - 3 \right) \bigg)^{-1}.
\end{align*}

\noindent
%where $\Sigma = -2 + \dfrac{3 \beta}{8 \pi} - \dfrac{5 K \beta}{8 \pi} \rho^{\Gamma-1}(r)$. 
%\\
%\\
Taking $\alpha = 1$, $\beta = 0$, and $\varphi = 0$ we recover the GR TOV equations.

\section{V. Numerical Modelling and Testing}

\noindent
In order to obtain a graphical relations of the TOV equations, we numerically integrate our derived TOV equations of the MIT bag model to this $f(T,\mathcal{T})$-gravity model for quark stars.
\\
\\
\noindent
We use the MIT bag model because it is the simplest equation of state for quark matter \cite{Kpadonou:2015eza, Astashenok:2014dja}. This is obtained because a quark star is a self-gravitating system consisting of deconfined $u$, $d$, and $s$ quarks and electrons \cite{Khoury:2003rn}. These deconfined quarks are the fundamental elements of the colour superconductor system \cite{Kpadonou:2015eza}. In the comparison with the standard hadron matter, they lead to a softer equation of state, the MIT bag model which is given in Eq. (\ref{MITbag}).
\\
\\
\noindent
The value of $\omega$ in Eq.(\ref{MITbag}) is dependent on the mass $m_{s}$ of the strange quark \cite{Kpadonou:2015eza}. In the case of radiation, we have $m_{s} = 0$ and the parameter is $\omega = 0$ \cite{Kpadonou:2015eza}. In the case of a more relativistic model having $m_{s} = 250$ MeV, the parameter would be $\omega = 0.28$ \cite{Astashenok:2014dja, Astashenok:2015qzw}. The parameter $\gamma_{0}$ lies within the intervals $58.8 < \gamma_{0} < 91.2$ which has units MeV/fm$^{3}$ \cite{Spergel:2003cb}.

\subsection{A. Mass Profile Curve}
\begin{center}
\begin{figure}[h!] 
\includegraphics[scale=0.325]{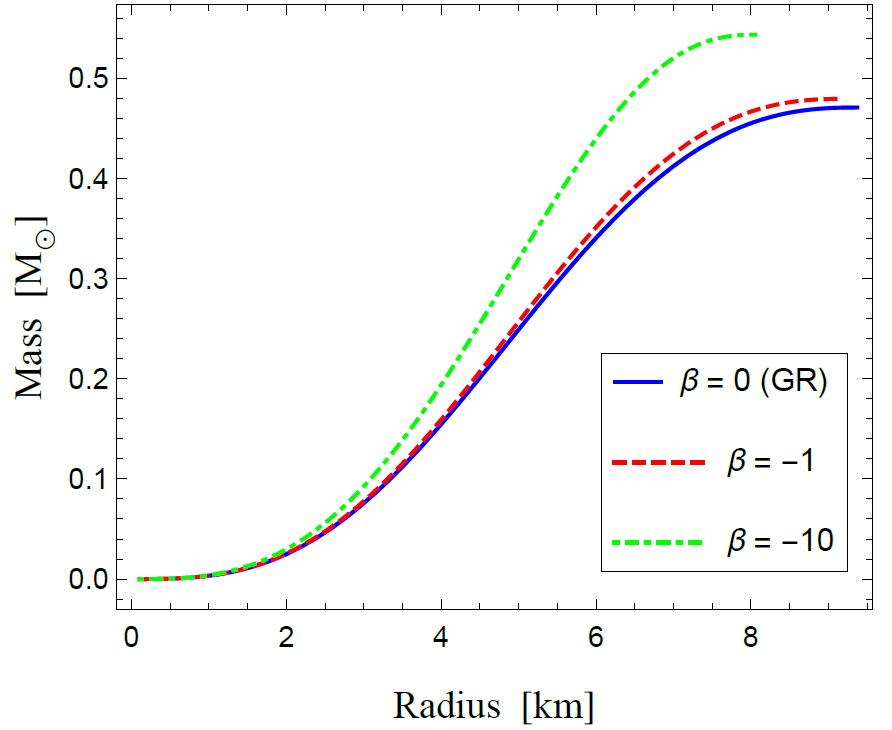}
\caption{Mass profile graph of a quark star obtained with $f = \alpha T(r) + \beta \mathcal{T}(r) + \varphi$ showing three different variations of $\beta$. The value of $\omega = 0.28$ and $\gamma_{0} = 1$} \label{figMR}
% \[Omega] = 0.28, Subscript[\[Gamma], o] = 1}
\end{figure}
\end{center}

\noindent
In Fig.(\ref{figMR}) we show the mass profile curve of a quark star by setting the values of $\alpha=1$, $\varphi=2.036 \times 10^{-35}$ (cosmological constant) \cite{carmeli2001value} then varying the value of $\beta$. 
\\
\\
We take three values of $\beta$ in this case to contrast between the GR case where $\beta = 0$, the case where the function for $\mathcal{T}(r) = \rho(r) - 3p(r)$ \cite{harko2014f} is included i.e. $\beta = -1$. Finally we include the case where this function is magnified by including $\beta = -10$ so as to see the behaviour at various levels. 
\\
\\
As we decrease the value of $\beta$ we allow for a smaller quark star structure. With the inclusion of the $\mathcal{T}(r)$ element some variations to arise. The quark star's maximum mass has however increased showing that having $\beta$ at lower orders of magnitude allows for a much denser quark star structure.
\\
\\
To show these variations properly we plot the curve for $\beta = -10$ in Fig.(\ref{figMR}). Here we note that for $\beta = -10$ a more massive quark star is allowed in such a gravity framework however smaller in size.

\subsection{B. Central Density-Radius Curve}

\begin{center}
\begin{figure}[h!] 
\includegraphics[scale=0.35]{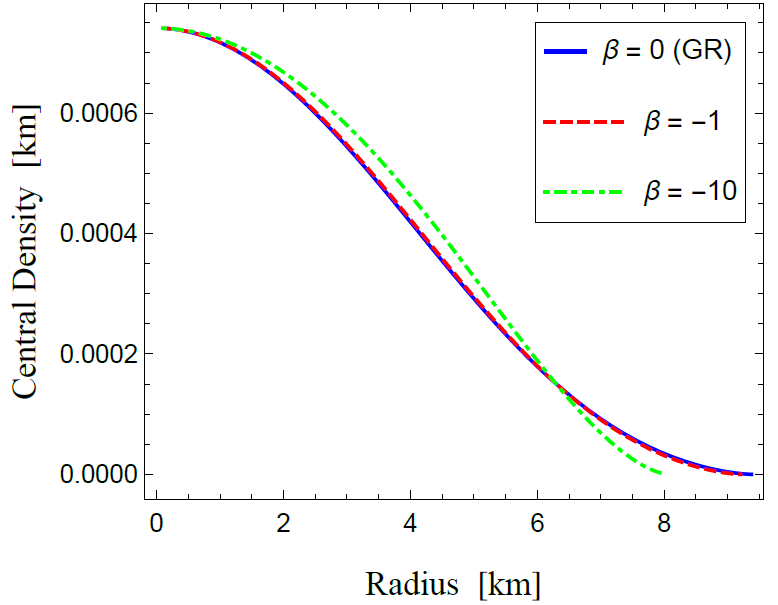}
\caption{Central Density-Radius graph of a quark star obtained with $f = \alpha T(r) + \beta \mathcal{T}(r) + \varphi$ showing three different variations of $\beta$. The value of $\omega = 0.28$ and $\gamma_{0} = 1$} \label{figDR}
% \[Omega] = 0.28, Subscript[\[Gamma], o] = 1}
\end{figure}
\end{center}

\noindent
In Fig.(\ref{figDR}) we also plot the central density-radius graph where we again set the values of $\alpha=1$, $\varphi=2.036 \times 10^{-35}$ (cosmological constant) \cite{carmeli2001value} then vary the value of $\beta$. 
\\
\\
Again we contrast with the GR case when taking $\beta = 0$. When we decrease the value of $\beta$ we may note that the central density figure of the quark star is more reluctant to drop however when reaching a certain radius it then decreases at a more rapid rate.
\\
\\
To further magnify this effect we again plot the results which are given by taking $\beta = -10$. In contrast to the GR case we see that the curve allows for a slightly denser quark star at a certain radius.

\section{VI. Conclusion}
\noindent
In this study we study the TOV equation and its derivative behaviour for the rotated, pure, spherically symmetric tetrad. We then contrasted this result to the GR case. Our model has responded well when the MIT bag model is considered.
\\
\\
Our main goal throughout this work was to keep our terms as general as possible, with the possibility to revert back to the GR case whenever we needed to. This fact was very useful in checking our results throughout the derivation.
\\
\\
Numerical techniques were required to solve the TOV equation where reasonable boundary conditions were used. We apply an equation of state so that we may eliminate one of the four variables i.e. make one of the variables dependent on another variable.
\\
\\
For future work we hope to be able to apply a lagrangian which is not linear however thus far we have been unable to yield working TOV equations.

\section{VII. Acknowledgements}
\noindent
\textit{The research work disclosed in this publication is funded by the ENDEAVOUR Scholarship Scheme (Malta). The scholarship may be part-financed by the European Union - European Social Fund (ESF) under Operational Programme II - Cohesion Policy 2014-2020, "Investing in human capital to create more opportunities and promote the well being of society"}.

%\begin{thebibliography}{200}
%\bibliography{Bibliography}

\begin{thebibliography}{38}
\expandafter\ifx\csname natexlab\endcsname\relax\def\natexlab#1{#1}\fi
\expandafter\ifx\csname bibnamefont\endcsname\relax
  \def\bibnamefont#1{#1}\fi
\expandafter\ifx\csname bibfnamefont\endcsname\relax
  \def\bibfnamefont#1{#1}\fi
\expandafter\ifx\csname citenamefont\endcsname\relax
  \def\citenamefont#1{#1}\fi
\expandafter\ifx\csname url\endcsname\relax
  \def\url#1{\texttt{#1}}\fi
\expandafter\ifx\csname urlprefix\endcsname\relax\def\urlprefix{URL }\fi
\providecommand{\bibinfo}[2]{#2}
\providecommand{\eprint}[2][]{\url{#2}}

\bibitem[{\citenamefont{Garnavich et~al.}(1998)\citenamefont{Garnavich, Jha,
  Challis, Clocchiatti, Diercks, Filippenko, Gilliland, Hogan, Kirshner,
  Leibundgut et~al.}}]{garnavich1998supernova}
\bibinfo{author}{\bibfnamefont{P.~M.} \bibnamefont{Garnavich}},
  \bibinfo{author}{\bibfnamefont{S.}~\bibnamefont{Jha}},
  \bibinfo{author}{\bibfnamefont{P.}~\bibnamefont{Challis}},
  \bibinfo{author}{\bibfnamefont{A.}~\bibnamefont{Clocchiatti}},
  \bibinfo{author}{\bibfnamefont{A.}~\bibnamefont{Diercks}},
  \bibinfo{author}{\bibfnamefont{A.~V.} \bibnamefont{Filippenko}},
  \bibinfo{author}{\bibfnamefont{R.~L.} \bibnamefont{Gilliland}},
  \bibinfo{author}{\bibfnamefont{C.~J.} \bibnamefont{Hogan}},
  \bibinfo{author}{\bibfnamefont{R.~P.} \bibnamefont{Kirshner}},
  \bibinfo{author}{\bibfnamefont{B.}~\bibnamefont{Leibundgut}},
  \bibnamefont{et~al.}, \bibinfo{journal}{The Astrophysical Journal}
  \textbf{\bibinfo{volume}{509}}, \bibinfo{pages}{74} (\bibinfo{year}{1998}).

\bibitem[{\citenamefont{Riess et~al.}(1998)\citenamefont{Riess, Filippenko,
  Challis, Clocchiatti, Diercks, Garnavich, Gilliland, Hogan, Jha, Kirshner
  et~al.}}]{riess1998observational}
\bibinfo{author}{\bibfnamefont{A.~G.} \bibnamefont{Riess}},
  \bibinfo{author}{\bibfnamefont{A.~V.} \bibnamefont{Filippenko}},
  \bibinfo{author}{\bibfnamefont{P.}~\bibnamefont{Challis}},
  \bibinfo{author}{\bibfnamefont{A.}~\bibnamefont{Clocchiatti}},
  \bibinfo{author}{\bibfnamefont{A.}~\bibnamefont{Diercks}},
  \bibinfo{author}{\bibfnamefont{P.~M.} \bibnamefont{Garnavich}},
  \bibinfo{author}{\bibfnamefont{R.~L.} \bibnamefont{Gilliland}},
  \bibinfo{author}{\bibfnamefont{C.~J.} \bibnamefont{Hogan}},
  \bibinfo{author}{\bibfnamefont{S.}~\bibnamefont{Jha}},
  \bibinfo{author}{\bibfnamefont{R.~P.} \bibnamefont{Kirshner}},
  \bibnamefont{et~al.}, \bibinfo{journal}{The Astronomical Journal}
  \textbf{\bibinfo{volume}{116}}, \bibinfo{pages}{1009} (\bibinfo{year}{1998}).

\bibitem[{\citenamefont{Copeland et~al.}(2006)\citenamefont{Copeland, Sami, and
  Tsujikawa}}]{copeland2006dynamics}
\bibinfo{author}{\bibfnamefont{E.~J.} \bibnamefont{Copeland}},
  \bibinfo{author}{\bibfnamefont{M.}~\bibnamefont{Sami}}, \bibnamefont{and}
  \bibinfo{author}{\bibfnamefont{S.}~\bibnamefont{Tsujikawa}},
  \bibinfo{journal}{International Journal of Modern Physics D}
  \textbf{\bibinfo{volume}{15}}, \bibinfo{pages}{1753} (\bibinfo{year}{2006}).

\bibitem[{\citenamefont{Bamba et~al.}(2012)\citenamefont{Bamba, Capozziello,
  Nojiri, and Odintsov}}]{bamba2012dark}
\bibinfo{author}{\bibfnamefont{K.}~\bibnamefont{Bamba}},
  \bibinfo{author}{\bibfnamefont{S.}~\bibnamefont{Capozziello}},
  \bibinfo{author}{\bibfnamefont{S.}~\bibnamefont{Nojiri}}, \bibnamefont{and}
  \bibinfo{author}{\bibfnamefont{S.~D.} \bibnamefont{Odintsov}},
  \bibinfo{journal}{Astrophysics and Space Science}
  \textbf{\bibinfo{volume}{342}}, \bibinfo{pages}{155} (\bibinfo{year}{2012}).

\bibitem[{\citenamefont{Feng et~al.}(2005)\citenamefont{Feng, Wang, and
  Zhang}}]{feng2005dark}
\bibinfo{author}{\bibfnamefont{B.}~\bibnamefont{Feng}},
  \bibinfo{author}{\bibfnamefont{X.}~\bibnamefont{Wang}}, \bibnamefont{and}
  \bibinfo{author}{\bibfnamefont{X.}~\bibnamefont{Zhang}},
  \bibinfo{journal}{Physics Letters B} \textbf{\bibinfo{volume}{607}},
  \bibinfo{pages}{35} (\bibinfo{year}{2005}).

\bibitem[{\citenamefont{Guo et~al.}(2005)\citenamefont{Guo, Piao, Zhang, and
  Zhang}}]{guo2005cosmological}
\bibinfo{author}{\bibfnamefont{Z.-K.} \bibnamefont{Guo}},
  \bibinfo{author}{\bibfnamefont{Y.-S.} \bibnamefont{Piao}},
  \bibinfo{author}{\bibfnamefont{X.}~\bibnamefont{Zhang}}, \bibnamefont{and}
  \bibinfo{author}{\bibfnamefont{Y.-Z.} \bibnamefont{Zhang}},
  \bibinfo{journal}{Physics Letters B} \textbf{\bibinfo{volume}{608}},
  \bibinfo{pages}{177} (\bibinfo{year}{2005}).

\bibitem[{\citenamefont{Boehmer et~al.}(2011)\citenamefont{Boehmer, Mussa, and
  Tamanini}}]{boehmer2011existence}
\bibinfo{author}{\bibfnamefont{C.~G.} \bibnamefont{Boehmer}},
  \bibinfo{author}{\bibfnamefont{A.}~\bibnamefont{Mussa}}, \bibnamefont{and}
  \bibinfo{author}{\bibfnamefont{N.}~\bibnamefont{Tamanini}},
  \bibinfo{journal}{Classical and Quantum Gravity}
  \textbf{\bibinfo{volume}{28}}, \bibinfo{pages}{245020}
  (\bibinfo{year}{2011}).

\bibitem[{\citenamefont{De~Felice and Tsujikawa}(2010)}]{de2010f}
\bibinfo{author}{\bibfnamefont{A.}~\bibnamefont{De~Felice}} \bibnamefont{and}
  \bibinfo{author}{\bibfnamefont{S.}~\bibnamefont{Tsujikawa}},
  \bibinfo{journal}{Living Rev. Rel} \textbf{\bibinfo{volume}{13}},
  \bibinfo{pages}{1002} (\bibinfo{year}{2010}).

\bibitem[{\citenamefont{Capozziello and
  De~Laurentis}(2011)}]{Capozziello:2011et}
\bibinfo{author}{\bibfnamefont{S.}~\bibnamefont{Capozziello}} \bibnamefont{and}
  \bibinfo{author}{\bibfnamefont{M.}~\bibnamefont{De~Laurentis}},
  \bibinfo{journal}{Phys. Rept.} \textbf{\bibinfo{volume}{509}},
  \bibinfo{pages}{167} (\bibinfo{year}{2011}), \eprint{1108.6266}.

\bibitem[{\citenamefont{Cai et~al.}(2016)\citenamefont{Cai, Capozziello,
  De~Laurentis, and Saridakis}}]{Cai:2015emx}
\bibinfo{author}{\bibfnamefont{Y.-F.} \bibnamefont{Cai}},
  \bibinfo{author}{\bibfnamefont{S.}~\bibnamefont{Capozziello}},
  \bibinfo{author}{\bibfnamefont{M.}~\bibnamefont{De~Laurentis}},
  \bibnamefont{and} \bibinfo{author}{\bibfnamefont{E.~N.}
  \bibnamefont{Saridakis}}, \bibinfo{journal}{Rept. Prog. Phys.}
  \textbf{\bibinfo{volume}{79}}, \bibinfo{pages}{106901}
  (\bibinfo{year}{2016}), \eprint{1511.07586}.

\bibitem[{\citenamefont{Capozziello et~al.}(2010)\citenamefont{Capozziello,
  De~Laurentis, and Faraoni}}]{Capozziello:2009nq}
\bibinfo{author}{\bibfnamefont{S.}~\bibnamefont{Capozziello}},
  \bibinfo{author}{\bibfnamefont{M.}~\bibnamefont{De~Laurentis}},
  \bibnamefont{and} \bibinfo{author}{\bibfnamefont{V.}~\bibnamefont{Faraoni}},
  \bibinfo{journal}{Open Astron. J.} \textbf{\bibinfo{volume}{3}},
  \bibinfo{pages}{49} (\bibinfo{year}{2010}), \eprint{0909.4672}.

\bibitem[{\citenamefont{Iorio and Saridakis}(2012)}]{iorio2012solar}
\bibinfo{author}{\bibfnamefont{L.}~\bibnamefont{Iorio}} \bibnamefont{and}
  \bibinfo{author}{\bibfnamefont{E.~N.} \bibnamefont{Saridakis}},
  \bibinfo{journal}{Monthly Notices of the Royal Astronomical Society}
  \textbf{\bibinfo{volume}{427}}, \bibinfo{pages}{1555} (\bibinfo{year}{2012}).

\bibitem[{\citenamefont{Unzicker and Case}(2005)}]{unzicker2005translation}
\bibinfo{author}{\bibfnamefont{A.}~\bibnamefont{Unzicker}} \bibnamefont{and}
  \bibinfo{author}{\bibfnamefont{T.}~\bibnamefont{Case}},
  \bibinfo{journal}{arXiv preprint physics/0503046}  (\bibinfo{year}{2005}).

\bibitem[{\citenamefont{Hayashi and Shirafuji}(1979)}]{hayashi1979new}
\bibinfo{author}{\bibfnamefont{K.}~\bibnamefont{Hayashi}} \bibnamefont{and}
  \bibinfo{author}{\bibfnamefont{T.}~\bibnamefont{Shirafuji}},
  \bibinfo{journal}{Physical Review D} \textbf{\bibinfo{volume}{19}},
  \bibinfo{pages}{3524} (\bibinfo{year}{1979}).

\bibitem[{\citenamefont{Tamanini and Boehmer}(2012)}]{Tamanini:2012hg}
\bibinfo{author}{\bibfnamefont{N.}~\bibnamefont{Tamanini}} \bibnamefont{and}
  \bibinfo{author}{\bibfnamefont{C.~G.} \bibnamefont{Boehmer}},
  \bibinfo{journal}{Phys. Rev.} \textbf{\bibinfo{volume}{D86}},
  \bibinfo{pages}{044009} (\bibinfo{year}{2012}), \eprint{1204.4593}.

\bibitem[{\citenamefont{Ade et~al.}(2014)}]{Planck:2013jfk}
\bibinfo{author}{\bibfnamefont{P.~A.~R.} \bibnamefont{Ade}}
  \bibnamefont{et~al.} (\bibinfo{collaboration}{Planck}),
  \bibinfo{journal}{Astron. Astrophys.} \textbf{\bibinfo{volume}{571}},
  \bibinfo{pages}{A22} (\bibinfo{year}{2014}), \eprint{1303.5082}.

\bibitem[{\citenamefont{Astashenok
  et~al.}(2015{\natexlab{a}})\citenamefont{Astashenok, Capozziello, and
  Odintsov}}]{astashenok2015nonperturbative}
\bibinfo{author}{\bibfnamefont{A.~V.} \bibnamefont{Astashenok}},
  \bibinfo{author}{\bibfnamefont{S.}~\bibnamefont{Capozziello}},
  \bibnamefont{and} \bibinfo{author}{\bibfnamefont{S.~D.}
  \bibnamefont{Odintsov}}, \bibinfo{journal}{Physics Letters B}
  \textbf{\bibinfo{volume}{742}}, \bibinfo{pages}{160}
  (\bibinfo{year}{2015}{\natexlab{a}}).

\bibitem[{\citenamefont{Farrugia et~al.}(2016)\citenamefont{Farrugia, Said, and
  Ruggiero}}]{farrugia2016solar}
\bibinfo{author}{\bibfnamefont{G.}~\bibnamefont{Farrugia}},
  \bibinfo{author}{\bibfnamefont{J.~L.} \bibnamefont{Said}}, \bibnamefont{and}
  \bibinfo{author}{\bibfnamefont{M.~L.} \bibnamefont{Ruggiero}},
  \bibinfo{journal}{Physical Review D} \textbf{\bibinfo{volume}{93}},
  \bibinfo{pages}{104034} (\bibinfo{year}{2016}).

\bibitem[{\citenamefont{Paliathanasis et~al.}(2016)\citenamefont{Paliathanasis,
  Barrow, and Leach}}]{paliathanasis2016cosmological}
\bibinfo{author}{\bibfnamefont{A.}~\bibnamefont{Paliathanasis}},
  \bibinfo{author}{\bibfnamefont{J.~D.} \bibnamefont{Barrow}},
  \bibnamefont{and} \bibinfo{author}{\bibfnamefont{P.}~\bibnamefont{Leach}},
  \bibinfo{journal}{arXiv preprint arXiv:1606.00659}  (\bibinfo{year}{2016}).

\bibitem[{\citenamefont{Krššák and Saridakis}(2016)}]{Krssak:2015oua}
\bibinfo{author}{\bibfnamefont{M.}~\bibnamefont{Krššák}} \bibnamefont{and}
  \bibinfo{author}{\bibfnamefont{E.~N.} \bibnamefont{Saridakis}},
  \bibinfo{journal}{Class. Quant. Grav.} \textbf{\bibinfo{volume}{33}},
  \bibinfo{pages}{115009} (\bibinfo{year}{2016}), \eprint{1510.08432}.

\bibitem[{\citenamefont{Harko et~al.}(2011)\citenamefont{Harko, Lobo, Nojiri,
  and Odintsov}}]{harko2011f}
\bibinfo{author}{\bibfnamefont{T.}~\bibnamefont{Harko}},
  \bibinfo{author}{\bibfnamefont{F.~S.} \bibnamefont{Lobo}},
  \bibinfo{author}{\bibfnamefont{S.}~\bibnamefont{Nojiri}}, \bibnamefont{and}
  \bibinfo{author}{\bibfnamefont{S.~D.} \bibnamefont{Odintsov}},
  \bibinfo{journal}{Physical Review D} \textbf{\bibinfo{volume}{84}},
  \bibinfo{pages}{024020} (\bibinfo{year}{2011}).

\bibitem[{\citenamefont{Nassur et~al.}(2015)\citenamefont{Nassur, Houndjo,
  Rodrigues, Kpadonou, and Tossa}}]{nassur2015early}
\bibinfo{author}{\bibfnamefont{S.}~\bibnamefont{Nassur}},
  \bibinfo{author}{\bibfnamefont{M.}~\bibnamefont{Houndjo}},
  \bibinfo{author}{\bibfnamefont{M.}~\bibnamefont{Rodrigues}},
  \bibinfo{author}{\bibfnamefont{A.}~\bibnamefont{Kpadonou}}, \bibnamefont{and}
  \bibinfo{author}{\bibfnamefont{J.}~\bibnamefont{Tossa}},
  \bibinfo{journal}{Astrophysics and Space Science}
  \textbf{\bibinfo{volume}{360}}, \bibinfo{pages}{1} (\bibinfo{year}{2015}).

\bibitem[{\citenamefont{Harko et~al.}(2014)\citenamefont{Harko, Lobo, Otalora,
  and Saridakis}}]{harko2014f}
\bibinfo{author}{\bibfnamefont{T.}~\bibnamefont{Harko}},
  \bibinfo{author}{\bibfnamefont{F.~S.} \bibnamefont{Lobo}},
  \bibinfo{author}{\bibfnamefont{G.}~\bibnamefont{Otalora}}, \bibnamefont{and}
  \bibinfo{author}{\bibfnamefont{E.~N.} \bibnamefont{Saridakis}},
  \bibinfo{journal}{Journal of Cosmology and Astroparticle Physics}
  \textbf{\bibinfo{volume}{2014}}, \bibinfo{pages}{021} (\bibinfo{year}{2014}).

\bibitem[{\citenamefont{Bengochea and Ferraro}(2009)}]{Bengochea:2008gz}
\bibinfo{author}{\bibfnamefont{G.~R.} \bibnamefont{Bengochea}}
  \bibnamefont{and} \bibinfo{author}{\bibfnamefont{R.}~\bibnamefont{Ferraro}},
  \bibinfo{journal}{Phys. Rev.} \textbf{\bibinfo{volume}{D79}},
  \bibinfo{pages}{124019} (\bibinfo{year}{2009}), \eprint{0812.1205}.

\bibitem[{\citenamefont{Ferraro and Fiorini}(2007)}]{Ferraro:2006jd}
\bibinfo{author}{\bibfnamefont{R.}~\bibnamefont{Ferraro}} \bibnamefont{and}
  \bibinfo{author}{\bibfnamefont{F.}~\bibnamefont{Fiorini}},
  \bibinfo{journal}{Phys. Rev.} \textbf{\bibinfo{volume}{D75}},
  \bibinfo{pages}{084031} (\bibinfo{year}{2007}), \eprint{gr-qc/0610067}.

\bibitem[{\citenamefont{Ferraro and Fiorini}(2008)}]{Ferraro:2008ey}
\bibinfo{author}{\bibfnamefont{R.}~\bibnamefont{Ferraro}} \bibnamefont{and}
  \bibinfo{author}{\bibfnamefont{F.}~\bibnamefont{Fiorini}},
  \bibinfo{journal}{Phys. Rev.} \textbf{\bibinfo{volume}{D78}},
  \bibinfo{pages}{124019} (\bibinfo{year}{2008}), \eprint{0812.1981}.

\bibitem[{\citenamefont{Linder}(2010)}]{Linder:2010py}
\bibinfo{author}{\bibfnamefont{E.~V.} \bibnamefont{Linder}},
  \bibinfo{journal}{Phys. Rev.} \textbf{\bibinfo{volume}{D81}},
  \bibinfo{pages}{127301} (\bibinfo{year}{2010}), \bibinfo{note}{[Erratum:
  Phys. Rev.D82,109902(2010)]}, \eprint{1005.3039}.

\bibitem[{\citenamefont{Deliduman and Yapiskan}(2011)}]{deliduman2011absence}
\bibinfo{author}{\bibfnamefont{C.}~\bibnamefont{Deliduman}} \bibnamefont{and}
  \bibinfo{author}{\bibfnamefont{B.}~\bibnamefont{Yapiskan}},
  \bibinfo{journal}{arXiv preprint arXiv:1103.2225}  (\bibinfo{year}{2011}).

\bibitem[{\citenamefont{Faraoni}(2000)}]{Faraoni:2000wk}
\bibinfo{author}{\bibfnamefont{V.}~\bibnamefont{Faraoni}},
  \bibinfo{journal}{Phys. Rev.} \textbf{\bibinfo{volume}{D62}},
  \bibinfo{pages}{023504} (\bibinfo{year}{2000}), \eprint{gr-qc/0002091}.

\bibitem[{\citenamefont{Paliathanasis et~al.}(2014)\citenamefont{Paliathanasis,
  Basilakos, Saridakis, Capozziello, Atazadeh, Darabi, and
  Tsamparlis}}]{Paliathanasis:2014iva}
\bibinfo{author}{\bibfnamefont{A.}~\bibnamefont{Paliathanasis}},
  \bibinfo{author}{\bibfnamefont{S.}~\bibnamefont{Basilakos}},
  \bibinfo{author}{\bibfnamefont{E.~N.} \bibnamefont{Saridakis}},
  \bibinfo{author}{\bibfnamefont{S.}~\bibnamefont{Capozziello}},
  \bibinfo{author}{\bibfnamefont{K.}~\bibnamefont{Atazadeh}},
  \bibinfo{author}{\bibfnamefont{F.}~\bibnamefont{Darabi}}, \bibnamefont{and}
  \bibinfo{author}{\bibfnamefont{M.}~\bibnamefont{Tsamparlis}},
  \bibinfo{journal}{Phys. Rev.} \textbf{\bibinfo{volume}{D89}},
  \bibinfo{pages}{104042} (\bibinfo{year}{2014}), \eprint{1402.5935}.

\bibitem[{\citenamefont{Kpadonou et~al.}(2016)\citenamefont{Kpadonou, Houndjo,
  and Rodrigues}}]{Kpadonou:2015eza}
\bibinfo{author}{\bibfnamefont{A.~V.} \bibnamefont{Kpadonou}},
  \bibinfo{author}{\bibfnamefont{M.~J.~S.} \bibnamefont{Houndjo}},
  \bibnamefont{and} \bibinfo{author}{\bibfnamefont{M.~E.}
  \bibnamefont{Rodrigues}}, \bibinfo{journal}{Astrophys. Space Sci.}
  \textbf{\bibinfo{volume}{361}}, \bibinfo{pages}{244} (\bibinfo{year}{2016}),
  \eprint{1509.08771}.

\bibitem[{\citenamefont{Astashenok et~al.}(2013)\citenamefont{Astashenok,
  Capozziello, and Odintsov}}]{astashenok2013further}
\bibinfo{author}{\bibfnamefont{A.~V.} \bibnamefont{Astashenok}},
  \bibinfo{author}{\bibfnamefont{S.}~\bibnamefont{Capozziello}},
  \bibnamefont{and} \bibinfo{author}{\bibfnamefont{S.~D.}
  \bibnamefont{Odintsov}}, \bibinfo{journal}{Journal of Cosmology and
  Astroparticle Physics} \textbf{\bibinfo{volume}{2013}}, \bibinfo{pages}{040}
  (\bibinfo{year}{2013}).

\bibitem[{\citenamefont{Astashenok
  et~al.}(2015{\natexlab{b}})\citenamefont{Astashenok, Capozziello, and
  Odintsov}}]{astashenok2015extreme}
\bibinfo{author}{\bibfnamefont{A.~V.} \bibnamefont{Astashenok}},
  \bibinfo{author}{\bibfnamefont{S.}~\bibnamefont{Capozziello}},
  \bibnamefont{and} \bibinfo{author}{\bibfnamefont{S.~D.}
  \bibnamefont{Odintsov}}, \bibinfo{journal}{Journal of Cosmology and
  Astroparticle Physics} \textbf{\bibinfo{volume}{2015}}, \bibinfo{pages}{001}
  (\bibinfo{year}{2015}{\natexlab{b}}).

\bibitem[{\citenamefont{Astashenok
  et~al.}(2015{\natexlab{c}})\citenamefont{Astashenok, Capozziello, and
  Odintsov}}]{Astashenok:2014dja}
\bibinfo{author}{\bibfnamefont{A.~V.} \bibnamefont{Astashenok}},
  \bibinfo{author}{\bibfnamefont{S.}~\bibnamefont{Capozziello}},
  \bibnamefont{and} \bibinfo{author}{\bibfnamefont{S.~D.}
  \bibnamefont{Odintsov}}, \bibinfo{journal}{Phys. Lett.}
  \textbf{\bibinfo{volume}{B742}}, \bibinfo{pages}{160}
  (\bibinfo{year}{2015}{\natexlab{c}}), \eprint{1412.5453}.

\bibitem[{\citenamefont{Khoury and Weltman}(2004)}]{Khoury:2003rn}
\bibinfo{author}{\bibfnamefont{J.}~\bibnamefont{Khoury}} \bibnamefont{and}
  \bibinfo{author}{\bibfnamefont{A.}~\bibnamefont{Weltman}},
  \bibinfo{journal}{Phys. Rev.} \textbf{\bibinfo{volume}{D69}},
  \bibinfo{pages}{044026} (\bibinfo{year}{2004}), \eprint{astro-ph/0309411}.

\bibitem[{\citenamefont{Astashenok and Odintsov}(2016)}]{Astashenok:2015qzw}
\bibinfo{author}{\bibfnamefont{A.~V.} \bibnamefont{Astashenok}}
  \bibnamefont{and} \bibinfo{author}{\bibfnamefont{S.~D.}
  \bibnamefont{Odintsov}}, \bibinfo{journal}{Phys. Rev.}
  \textbf{\bibinfo{volume}{D94}}, \bibinfo{pages}{063008}
  (\bibinfo{year}{2016}), \eprint{1512.07279}.

\bibitem[{\citenamefont{Spergel et~al.}(2003)}]{Spergel:2003cb}
\bibinfo{author}{\bibfnamefont{D.~N.} \bibnamefont{Spergel}}
  \bibnamefont{et~al.} (\bibinfo{collaboration}{WMAP}),
  \bibinfo{journal}{Astrophys. J. Suppl.} \textbf{\bibinfo{volume}{148}},
  \bibinfo{pages}{175} (\bibinfo{year}{2003}), \eprint{astro-ph/0302209}.

\bibitem[{\citenamefont{Carmeli and Kuzmenko}(2001)}]{carmeli2001value}
\bibinfo{author}{\bibfnamefont{M.}~\bibnamefont{Carmeli}} \bibnamefont{and}
  \bibinfo{author}{\bibfnamefont{T.}~\bibnamefont{Kuzmenko}},
  \bibinfo{journal}{arXiv preprint astro-ph/0102033}  (\bibinfo{year}{2001}).

\end{thebibliography}
%end{thebibliography}

\end{document}